# Power-Flow-Embedded Projection Conic Matrix Completion for Low-Observable Distribution Systems

Xuzhuo Wang, *Student Member, IEEE*, Guoan Yan, *Student Member, IEEE*, Zhengshuo Li, *Senior Member, IEEE*

*Abstract*—A low-observable distribution system has insufficient measurements for conventional weighted least square state estimators. Matrix completion state estimators have been suggested, but their computational times could be prohibitive. To resolve this problem, a novel and efficient power-flow-embedded projection conic matrix completion method customized for low-observable distribution systems is proposed in this letter. This method can yield more accurate state estimations (2-fold improvement) in a much shorter time (5% or less) than other methods. Case studies on different-scale systems have demonstrated the efficacy of the proposed method when applied to low-observable distribution system state estimation problems.

## I. INTRODUCTION

DISTRIBUTION systems (DSs) under low-observable conditions lack enough measurements for conventional weighted least square state estimators to function normally [1]. Conventional remedies to this problem include installing additional sensors or leveraging pseudo-measurements [2],[3]. In addition to these remedies, based on semidefinite programming (SDP) in [4], a matrix completion state estimation (MCSE) method was proposed in [5] that was able to estimate bus voltages even under low-observable conditions where traditional state estimators may fail. However, this type of MCSE method may not always yield satisfying results; more importantly, its computational time could be prohibitive.

To solve this problem, this work improves upon the MCSE method in two aspects. First, in terms of **accuracy**, inspired by the very recent work [6], we propose a novel power-flow-embedded projection conic matrix completion model for the low-observable DS state estimation problem. Compared with the general-purpose model in [6], linearized power flow constraints are embedded in our model that can be regarded as a sort of "*domain knowledge*", which further increases the state estimators' accuracy. Second, in terms of **computational time**, two acceleration solution strategies, based on the sparse positive semidefinite approximation and a tailored branch-and-bound algorithm respectively, are proposed to solve the aforementioned model in a much shorter time. Case studies have demonstrated that, compared to the existing MCSE method presented in [5], our method exhibits an approximately 2-fold improvement in accuracy, and the computational time can be reduced to as short as 5% or even less.

## II. PRELIMINARIES: CONVENTIONAL MCSE IN [5]

Given a low-rank matrix $M \in \mathbb{R}^{n \times m}$ for which only a portion of the entries are known, [4] proposed recovering the unknown entries of the matrix by solving the following matrix completion problem:

This work was supported by National Key R&D Program of China under grant 2022YFB2402900. X. Wang, G. Yan and Z. Li are with Shandong University, Jinan 250061, China. X. Wang and G. Yan contributed equally and should be considered co-first authors. Zhengshuo Li is the corresponding author (e-mail: zsli@sdu.edu.cn).

$$\min_{D_1,D_2,X} tr(D_1) + tr(D_2) \tag{1a}$$

$$\text{s. t. } X_{ij} = M_{ij} \ \forall (i,j) \in \psi \tag{1b}$$

$$\begin{bmatrix} D_1 & X \\ X^* & D_2 \end{bmatrix} \succeq 0 \tag{1c}$$

where $\psi$ represents the set of subscripts of the known elements; $tr(\cdot)$ represents the trace of a matrix; the superscript * denotes the conjugate transpose of a matrix; and among the matrix variables $X$, $D_1$ and $D_2$, $X$ is the "recovered" version of $M$.

In [5], the above approach was leveraged to estimate the state variables in a low-observable DS. Specifically, an $n \times m$ measurement matrix, denoted as $M$, is first constructed, where each row represents a bus and each column represents a measurement quantity relevant to that bus. Here, "$n$" is the number of buses, and "$m$" is the number of measurements. For instance, if $m = 5$, each row can be formulated as $[\text{Re}(v_i), \text{Im}(v_i), |v_i|, \text{Re}(s_i), \text{Im}(s_i)], (1 \leq i \leq n)$ [5]. In this formulation, since matrix $M$ contains the real and imaginary parts of the bus voltages, one can straightforwardly obtain an estimation of the bus voltage phasors after $M$ is recovered. Hence, in [5], this method was regarded as a special *state estimator*, namely, the MCSE.

In [5], it was assumed that only a few elements in the measurement matrix $M$ are known due to the low observability and that $M$ can be generally regarded as a low-rank matrix (that is, $rank(M) < \min(n, m)$). Therefore, the missing elements in $M$ can be estimated by model (1). Furthermore, [5] noted that embedding the power flow constraints into (1) can increase the accuracy of the voltage phasor estimation and suggested modifying the equality constraint (1b) into (2c) to address the measurement noise. The MCSE model in [5] is thus given as follows:

$$\min_{D_1,D_2,X} tr(D_1) + tr(D_2) + \sum_{\varepsilon \in E} \omega_\varepsilon \mathcal{L}(\varepsilon) \tag{2a}$$

$$\text{s. t. (1c)} \tag{2b}$$

$$\|X_\psi - M_\psi\|_F \leq \delta \tag{2c}$$

$$\left. \begin{array}{c} \begin{bmatrix} -\tau_{\text{Re}} \\ -\tau_{\text{Im}} \end{bmatrix} \leq \begin{bmatrix} \text{Re}(\mathbf{v} - \mathbf{A}\begin{bmatrix} \text{Re}(\mathbf{s}) \\ \text{Im}(\mathbf{s}) \end{bmatrix} - \mathbf{w}) \\ \text{Im}(\mathbf{v} - \mathbf{A}\begin{bmatrix} \text{Re}(\mathbf{s}) \\ \text{Im}(\mathbf{s}) \end{bmatrix} - \mathbf{w}) \end{bmatrix} \leq \begin{bmatrix} \tau_{\text{Re}} \\ \tau_{\text{Im}} \end{bmatrix} \\ -\gamma \leq |\mathbf{v}| - \left( \mathbf{C}\begin{bmatrix} \text{Re}(\mathbf{s}) \\ \text{Im}(\mathbf{s}) \end{bmatrix} + |\mathbf{w}| \right) \leq \gamma \\ \begin{bmatrix} -\alpha_{\text{Re}} \\ -\alpha_{\text{Im}} \end{bmatrix} \leq \begin{bmatrix} \text{Re}(\mathbf{s}_0 - (\mathbf{v}_0(\overline{\mathbf{Y}}_{00}\overline{\mathbf{v}}_0 + \overline{\mathbf{Y}}_{0L}\overline{\mathbf{v}}))) \\ \text{Im}(\mathbf{s}_0 - (\mathbf{v}_0(\overline{\mathbf{Y}}_{00}\overline{\mathbf{v}}_0 + \overline{\mathbf{Y}}_{0L}\overline{\mathbf{v}}))) \end{bmatrix} \leq \begin{bmatrix} \alpha_{\text{Re}} \\ \alpha_{\text{Im}} \end{bmatrix} \end{array} \right\} \tag{2d}$$

where the abbreviations of the bus admittance matrices $\mathbf{Y}_{00}$ and $\mathbf{Y}_{0L}$, $\mathbf{w}$ as the vector of non-slack zero-load voltages, and the constant complex matrices $\mathbf{A}$ and $\mathbf{C}$, are referred to in [5]; the vectors $\mathbf{v}_0$, $\mathbf{s}_0$ collect the slack bus voltage phasor and power injection; the vectors $\mathbf{v}$ and $\mathbf{s}$ collect non-slack bus voltages and power injections; $\mathcal{L}(\cdot)$ denotes a loss function, e.g., the absolute value for the $\ell_1$ loss and the square for the squared-$\ell_2$ loss; $\|\cdot\|_F$ represents the Frobenius norm; Re(·), Im(·) and $|\cdot|$ represent the real part, imaginary part and absolute value of vectors, respectively; the set $\mathbf{E}=\{\tau_{Re},\tau_{Im},\gamma,\alpha_{Re},\alpha_{Im}\}$ collects the error tolerances; $\omega_\varepsilon$ is the associated weight of each constraint tolerance $\varepsilon \in \mathbf{E}$; and $\delta \geq 0$ is a parameter that can be tuned based on the extent of measurement noise. (2d) shows the Cartesian linearization of the exact AC power flow equations [7].

As mentioned above, after solving this model (2), the bus phasors and the other missing measurements can be obtained.

## III. Proposed Method

### A. Power-Flow-Embedded Projection Conic Model

Recently, in [6], a new projection conic model was proposed for general-purpose matrix completion as follows:

$$\min_{X,\Upsilon,\Theta,U} \|X_\psi - M_\psi\|_F + \lambda tr(\Theta) \tag{3a}$$

$$\text{s. t.} \begin{bmatrix} \Upsilon & X \\ X^T & \Theta \end{bmatrix} \succeq 0 \tag{3b}$$

$$\begin{bmatrix} \Upsilon & U \\ U^T & I_k \end{bmatrix} \succeq 0 \tag{3c}$$

$$\mathbf{0} \preceq \Upsilon \preceq I_n, \; tr(\Upsilon) \leq k \tag{3d}$$

where $U \in \mathbb{R}^{n \times k}$, $\Upsilon \in \mathbb{R}^{n \times n}$ and $\Theta \in \mathbb{R}^{m \times m}$; the orthogonal projection matrix $\Upsilon$, which satisfies $\Upsilon^2 = \Upsilon$, is relaxed to (3c) and (3d); $\mathbf{0}$ denotes the zero matrix; $I_k$ and $I_n$ denote the $k$-dimensional and $n$-dimensional identity matrix; $\lambda > 0$ is a parameter that regularizes $X$ to control its sensitivity to noise; and $k$ is a hyperparameter that bounds the rank of $X$, which, according to [6], can be specified as a value in {1,2,3,4,5}. However, in our target state estimator application, different values of $k \in \{1,2,3,4,5\}$ will lead to a difference of only approximately 0.1% between the mean absolute percent errors (MAPEs) of the solutions. Therefore, $k$ can be set to an arbitrary value in {1,2,3,4,5} for the state estimator problems in Section IV.

In comparison with model (1) and model (2), model (3) introduces the projection matrix $\Upsilon$ and the associated auxiliary matrix variables $U$ and $\Theta$, which, as shown in [6], will yield better solutions for a general class of matrix completion problems. The theoretical explanation can be found in [6] and is thus omitted.

Nevertheless, model (3) is general-purpose and does not consider the *"domain knowledge"* in DSs, e.g., the linearized power flow relationships in (2d). As shown in [5], embedding such relationships can extend a general matrix completion model into the low-observable DSs problem to obtain more accurate estimations. Hence, we further embed (2d) into (3) to obtain the following power-flow-embedded projection conic matrix completion model customized for the state estimation application:

$$\min_{X,\Upsilon,\Theta,U} \|X_\psi - M_\psi\|_F + \lambda tr(\Theta) + \sum_{\varepsilon \in \mathbf{E}} \omega_\varepsilon \mathcal{L}(\varepsilon) \tag{4a}$$

$$\text{s. t. (2d) and (3b)-(3d)} \tag{4b}$$

While the accuracy can be improved by the above modeling proposal, another significant problem must also be resolved: how to efficiently solve this model (4). Since the computational time of a direct solution by calling off-the-shelf SDP solvers can be prohibitive, two solution strategies, S1 and S2, are proposed below.

### B. Strategy S1: Sparse Positive Semidefinite Approximation

To pursue a more efficient solution but at the cost of an allowable accuracy loss, one needs to overcome the challenge of solving the large-scale SDP problem (4). A natural method for doing so involves utilizing the fact that a positive semidefinite symmetric matrix implies all its principal minors are positive semidefinite. First, the constraint in (3b) is rewritten as follows:

$$W = \begin{bmatrix} \Upsilon & X \\ X^T & \Theta \end{bmatrix} \in S_+^{m+n} \tag{5}$$

where $S_+^{m+n}$ denotes the cone of $(m+n) \times (m+n)$ symmetric positive semidefinite matrices.

Then, based on the so-called sparse positive semidefinite approximation approach [8], constraint (5) can be replaced or relaxed by enforcing positive semidefinite constraints on all or some smaller $d \times d$ principal submatrices of $W$, i.e.,

$$W_i \in S_+^d, \forall i=1,...,n_d \tag{6}$$

where $n_d$ is the number of selected $d \times d$ ($2 \leq d \leq n$) primary submatrices of $W \in S_+^{m+n}$.

Apparently, replacing (5) with a series of positive semidefinite problem on smaller $d \times d$ principal submatrices in (6) is likely to greatly reduce the solution time. The remaining question is how the hyperparameters $d$ and $n_d$ affect the computational performance of S1, which is tested in Section IV.

### C. Strategy S2: Tailored Branch-and-Bound Algorithm

In [6], a branch-and-bound algorithm is proposed for solving model (3), the general idea of which is illustrated in Fig. 1. In each iteration, problem (7) in the gray box is solved. Model (7) is as follows:

$$\min_{X,\Upsilon,\Theta,U} \|X_\psi - M_\psi\|_F + \lambda tr(\Theta) \tag{7a}$$

$$\text{s. t. (3b)-(3d)} \tag{7b}$$

$$\|U_i^T z\|_2^2 \leq \begin{cases} z^T \hat{U}_i \hat{U}_i^T z + (\hat{U}_i - U_i)^T z & \text{if } U_i^T z \in [\underline{b}_i, \hat{U}_i^T z] \\ z^T \hat{U}_i \hat{U}_i^T z + (U_i - \hat{U}_i)^T z & \text{if } U_i^T z \in [\hat{U}_i^T z, \overline{b}_i] \end{cases}, \forall i \in [k] \tag{7c}$$

where $\hat{U}_i$ is the $i$-th column of $\hat{U}$, $z$ is the eigenvector of $\hat{\Upsilon} - \hat{U}\hat{U}^T$, $\underline{b}_i$ and $\overline{b}_i$ are the constraint limits of (7c) for this new branch, restricting the feasible region of $U_i$. All of the above variables are updated with each iteration.

After problem (7) is solved, by checking this solution, the algorithm will either terminate or go to a branch-and-bound process, which prepares it for the next round of iteration. It should be noted that to save space, Fig. 1 presents a schematic flowchart only; the detailed branch-and-bound process can be found in [6].

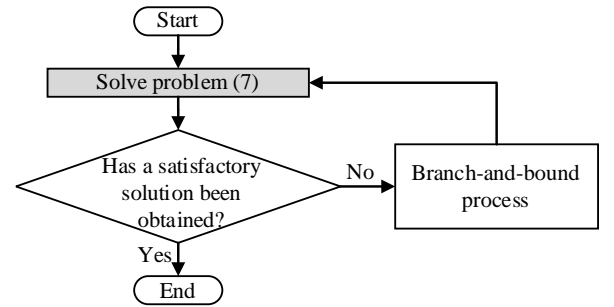

Fig. 1. Schematic flowchart of the branch-and-bound algorithm in [6].

In principle, this original branch-and-bound algorithm is applicable to (4). Nevertheless, for relatively large-scale DSs, the computational time of this algorithm remains too long due to the presence of a large-scale SDP matrix in the model in the gray box in Fig. 1. To address this issue, we tailor the original algorithm by solving the following problem (8) instead of (7) in the gray box in Fig. 1 for each iteration. This is due to the similar reasons explained in Section III.B, and this tailored version will be faster than the original version in [6].

$$\min_{X,\Upsilon,\Theta,U} \|X_\psi - M_\psi\|_F + \lambda tr(\Theta) + \sum_{\varepsilon \in E} \omega_\varepsilon \mathcal{L}(\varepsilon) \quad (8a)$$

$$\text{s. t. (2d), (3c), (3d), (6) and (7c)} \quad (8b)$$

## IV. CASE STUDIES

### A. Validation of the Efficacy of the Proposed Method

The conventional MCSE model (2) (**M1**) in [5] and the general-purpose model (3) (**M2**) in [6] are compared with the solutions of our model (4) that are solved by Strategies **S1** and **S2**. Details of the parameter values of M1 and M2 are provided in [5] and [6]. Tests are conducted on 141-bus [9] and 533-bus DSs [10]. Following [11], the low-observability condition is measured by the index of the fraction of the available data (FAD), which is set to 0.32 here; the specific measurement selection method can be found in that study. The generation method of known measurement data and measurement errors is the same as that in [11]. The MAPEs of the estimated voltage magnitude is chosen as the index of accuracy. All the SDP models appearing in M1, M2, S1 and S2 in these tests are solved using MATLAB/MOSEK.

TABLE I
ACCURACY IN TERMS OF MEAN ABSOLUTE PERCENT ERRORS (MAPES) (UNIT: %)

| MAPE | M1 | M2 | S1 | S2 |
|---|---|---|---|---|
| 141-bus system | 1.07 | >10 | 0.56 | 0.42 |
| 533-bus system | N/A | N/A | 0.54 | 0.50 |

TABLE II
COMPUTATIONAL TIME (UNIT: SECOND)

| Time | M1 | M2 | S1 | S2 |
|---|---|---|---|---|
| 141-bus system | 42.3 | 53.8 | 1.7 | 38.3 |
| 533-bus system | >1800 | >1800 | 4.3 | 262.1 |

In this array of tests, where $k=1$, $\lambda=10$, and $d=5$ for both systems, we have $n_d = 700$ for the 141-bus system and $n_d = 2000$ for the 533-bus system. Tables I and II show the estimation accuracy and computational time, respectively, and N/A indicates that there is no solution for that test. It can be seen that (i) compared to M1 and M2, both S1 and S2 have better accuracy (approximately 2-fold improvement over M1) and shorter computational times, while the M1 and M2 methods cannot yield solutions within 30 minutes for the 533-bus system; (ii) moreover, compared to S2, S1 has a slight accuracy loss on the order of 0.1%, but its computational time is much shorter and even less than $1.7/42.3 \approx 5\%$ of that of M1 and M2; (iii) S2, which solves model (4) embedded with power flow constraints, is significantly more accurate than M2, indicating that "*domain knowledge*" of DSs is indispensable when addressing low-observable DS problems. Incidentally, unlike our tailored version, i.e., S2, the original branch-and-bound algorithm in [6] yields no solution within 30 minutes for the 533-bus system.

### B. Influence of Hyperparameters

There are two hyperparameters $d$ and $n_d$ in strategy S1, and the impacts of different combinations of their values on this strategy's computational time and accuracy in the 141-bus system are shown in Table III.

By comparing Tables I, II and III, it can be seen that (i) as the values of $d$ and $n_d$ increase, the computational time increases, but the computational time of S1 remains less than 2 seconds, significantly less than that of M1 and M2; (ii) any combination of $d \geq 5$ in Table III has better computational performance than M1 and M2 in Table I. This indicates that S1 has an advantage over M1 and reflects that S1 is *robust*, i.e., it does not depend on a carefully selected, or special combination of $d$ and $n_d$.

TABLE III
COMPUTATIONAL PERFORMANCE OF S1 WITH DIFFERENT VALUES OF $d$ AND $n_d$

| Performance Index | | $n_d = 500$ | $n_d = 700$ | $n_d = 900$ |
|---|---|---|---|---|
| MAPE (%) | $d = 3$ | 3.40 | 3.40 | 3.40 |
| | $d = 5$ | 0.61 | 0.60 | 0.57 |
| | $d = 7$ | 0.60 | 0.59 | 0.56 |
| Time (Second) | $d = 3$ | 0.11 | 0.15 | 1.12 |
| | $d = 5$ | 0.66 | 0.89 | 1.66 |
| | $d = 7$ | 0.70 | 1.17 | 1.70 |

## V. CONCLUSION

In this letter, a novel and efficient power-flow-embedded projection conic matrix completion method is proposed for low-observable DS state estimation problems. Two strategies are suggested and compared to the existing methods. The superiority of the proposed method in terms of both accuracy and computational time has been demonstrated. Although the efficacy of our method is demonstrated mainly as a low-observable DS state estimator, it is worth noting that our method is also valuable to a state estimator for an observable DS because it can be used to generate additional pseudo-measurements, increasing the redundancy of the estimator. Future work will include a theoretical investigation into the latent relationships between the hyperparameters in the proposed method.

## VI. REFERENCES

[1] A. Abur and A. G. Exposito, *Power System State Estimation: Theory and Implementation*. Boca Raton, FL, USA: CRC Press, 2004.
[2] S. Bhela, V. Kekatos, and S. Veeramachaneni, "Enhancing observability in distribution grids using smart meter data," *IEEE Trans. Smart Grid*, vol. 9, no. 6, pp. 5953–5961, Nov. 2018.
[3] E. Manitsas, R. Singh, B. C. Pal, and G. Strbac, "Distribution system state estimation using an artificial neural network approach for pseudo measurement modeling," *IEEE Trans. Power Syst.*, vol. 27, no. 4, pp. 1888–1896, Nov. 2012.
[4] E. J. Candès, B. Recht, "Exact matrix completion via convex optimization," *Found. Comput. Math.*, vol. 9, no. 6, pp. 717–772, Apr. 2009.
[5] P. L. Donti, Y. Liu, A. J. Schmitt, A. Bernstein, R. Yang and Y. Zhang, "Matrix completion for low-observability voltage estimation," *IEEE Trans. Smart Grid.*, vol. 11, no. 3, pp. 2520-2530, May. 2020.
[6] D. Bertsimas *et al*., "Optimal low-rank matrix completion: semidefinite relaxations and eigenvector disjunctions," *arXiv: 2305.12292*, 2023.
[7] A. Bernstein and E. Dall'Anese, "Linear power-flow models in multiphase distribution networks," *Proc. IEEE PES Innov. Smart Grid Technol. Conf. Europe (ISGT-Europe)*, pp. 1–6, Sep. 2017.
[8] G. Blekherman *et al*., "Sparse PSD approximation of the PSD cone," *Math. Program.,* vol.191, pp.981–1004, 2022.
[9] R. D. Zimmerman, C. E. Murillo-Sánchez, and R. J. Thomas, "MATPOWER: Steady-state operations, planning and analysis tools for power systems research and education", *IEEE Trans. Power Syst.*, vol.26, no.1, pp.12–19, Feb. 2011.
[10] M. Gabriel and T. Lovisa, "Network reconfiguration for renewable generation maximization," M.S. thesis, LTH, Lund Univ., Lund, Sweden, 2023.
[11] B. Liu, H. Wu, Y. Zhang, R. Yang and A. Bernstein, "Robust matrix completion state estimation in distribution systems," *IEEE Power & Energy Society General Meeting (PESGM)*, Atlanta, GA, USA, pp. 1-5, 2019.